\documentclass[10pt,a4paper]{article}
\usepackage[utf8]{inputenc}
\usepackage{authblk}
\usepackage{xcolor}
\usepackage{mathtools, cuted}
\usepackage{amsmath}
\usepackage{amsfonts}
\usepackage{amssymb}
\usepackage{graphicx}
\usepackage{subcaption}
\usepackage[left=2cm, right=2cm, top=3cm, bottom=3cm]{geometry}
\title{The invariant based investigation of Shortcut to Adiabaticity for Quantum Harmonic Oscillators under time varying frictional force}
\author{T. Kiran and  M. Ponmurugan\footnote{email: ponphy@cutn.ac.in} }
\affil{Department of Physics, School of Basic and Applied Sciences, Central University of Tamilnadu, Thiruvarur - 610 005, Tamilnadu, India}
\date{}
\begin{document}
	\maketitle
	
	\subsection*{Abstract:} We investigate the Shortcuts To Adiabaticity (STA) of a quantum harmonic oscillator under time-dependent frictional force, using invariant based inverse engineering method with a class of invariants characterized by a time-dependent frictional coefficient. We discuss the implementation of shortcut protocol in a generalized framework for any arbitrary time-dependent frictional interaction and study a class of such interaction from which the STA for the quantum harmonic oscillator can be obtained. We also discuss the application of the above protocol for the harmonic oscillator with time-varying mass. For an illustration, we consider the coupled photonic lattice as a harmonic oscillator with time-varying mass and frequency and discuss implementing the above protocol.
	
	\begin{section}{Introduction}
		Shortcuts To Adiabaticity (STA) protocols are non-adiabatic processes that reproduce in finite time the same initial and final states as that of an infinitely slow adiabatic  process~\cite{Torrontegui,Guery2019,del2019}. These protocols can be used as an alternate driving of the system to implement the adiabatic process. The transition path will be different and decided by various factors involved in the specified technique of STA. Interestingly, there is no need for the complete suppression of the unwanted transition throughout the path. However, the initial and final states of the overall process need to be adiabatic in all sense. In other words, the STA process will mimic the dynamics of a prolonged adiabatic process within a finite time by allowing transitions at intermediate times~\cite{Boshier2012,Ibez2012prl,Mathieu,Deffner2014prx}. Experiments confirmed the feasibility of such processes on various grounds, noticeably for the frictionless transport of trapped ions~\cite{Bowler,Walther2012,Shuoming}, cold atoms~\cite{Shujin, Yan}, fast equilibration of a Brownian particle~\cite{Ignacio2016} and high-fidelity driving of a Bose-Einstein condensate~\cite{Mark2012}. Different kinds of methods are developed so far to establish the adiabaticity through non-adiabatic transitions. Some of them are Counterdiabatic Driving process by incorporating a global Hamiltonian to suppress the non-adiabatic transitions~\cite{Mustafa2003,Mustafa2005,Berry2009,Adolfo2012}, Local-Counterdiabatic Driving, where the local potential take charge of counterdiabatic contribution~\cite{Ibez2012prl}, Fast-forward approach~\cite{Shumpei2008,Shumpei2010}, and invariant based Inverse Engineering (IE) method by using the  Lewis-Riesenfeld (LR) invariants to connect the initial and final states through a non-adiabatic path~\cite{Lewis1969,Chen2010,Torrontegui2011,Chen2011pra,Chen2011,Torrontegui2012}. IE method is found useful in many applications and recently considered it in the context of the cost of shortcut process~\cite{Obinna2018,Obinna2019,baris2019}.
		
		The invariant method is employed extensively to find the solution for a given system under frictional contact~\cite{Khandekar1979}. The frictional force's time-dependency is also considered, which brings the concept of parametric variation of frictional force or equivalent change in the mass of the system. The mass varying Quantum Harmonic Oscillator (QHO) is considered in the context of STA using the IE method and connected it with photonic lattice~\cite{Stefanatos2014}. The mass varying QHO Hamiltonian is mathematically equivalent to the QHO experiencing a frictional force, which is a relevant topic since the works of Caldirola~\cite{Caldirola1941} and Kanai~\cite{Kanai1948}, discussing the idea of quantization of the systems experiencing certain types of non-conservative forces. Physical description of friction in these systems was debated for years, and still, these two interpretations are valid, one with dissipating energy and another with exponentially varying mass~\cite{Caldirola1941,Kanai1948,Baskoutas1993,Jeong2013}. We can generate a class of invariants by following the method in ref.~\cite{Lewis1969} for the Hamiltonian of a QHO experiencing time-dependent frictional force~\cite{Pedrosa1987,Pedrosa1997}. The class of invariants for the above system are characterized by the different solutions of Ermakov equation~\cite{John1981}. Among these invariants, a particular choice of invariant can be used to implement STA for QHO by using the time-dependent control of frictional force, which is not explored in the context of STA.  
		
		Recent developments of invariants for quantum dissipative systems referred to the LR invariant as the strong invariant to distinguish it from the weak invariant for open quantum systems~\cite{Sumiyoshi2016}. In this work, we follow the LR invariant formulation to develop the STA protocol since 
the system Hamiltonian completely describes the time evolution of the system, unlike, in the case of open system consideration. We develop a general framework of STA for QHO under time-dependent frictional force and study a class of such oscillators to achieve STA for ideal time-dependent QHO. We also explain the use of our framework in the context of QHO with time-dependent mass and applicability in coupled photonic lattices. As the invariant method found to be the most efficient way to implement STA in thermodynamic engines~\cite{Obinna2018}, studying oscillator under frictional force using invariant will be useful in quantum thermal engine studies. Also, the frictional force's arbitrary time-dependence can be used to improve the protocols to drive adiabatic strokes of quantum thermal engines.
		
		In section two of this paper, we investigate a class of invariants and corresponding Ermakov equation for the QHO under time-dependent frictional force. We also discuss the necessary boundary conditions to establish the generalized framework of STA for all arbitrary time dependency of frictional force in the third section. Following the general formalism, we identify a set of QHO under time-dependent frictional force in section four. It is possible to achieve STA for an ideal time-dependent QHO from that, and a feasible protocol is illustrated by choosing an appropriate solution for the Ermakov equation called the scaling factor. We analyze the characteristics of such an STA protocol, including the cost of implementation. In section five, we consider a QHO with time-dependent mass as an illustration to prove that the STA protocol developed for QHO under time-dependent frictional force can be applied for QHO with time-dependent mass. We use the same shortcut protocol to get the desired output in a photonic lattice described by the differential set similar to the time-dependent Schrodinger equation with mass varying QHO Hamiltonian. STA is achieved by arbitrarily controlling its lattice parameters as a function of propagation distance. Finally, we summarize our results in the conclusion section. 		
	\end{section}
	\begin{section}{Quantum harmonic oscillator with time varying friction}

			Consider the time-dependent QHO Hamiltonian~\cite{Pedrosa1987,Pedrosa1997},
		\begin{equation}
		\mathcal{\hat{H}}=f\left(t\right)\frac{\hat{p}^{2}}{2}+f^{-1}\left(t\right)\frac{\omega^{2}(t)\hat{x}^{2}}{2},
		\end{equation}
		whose angular frequency and function $f(t)$ explicitly depends on time. Using the commutation relation between the position and the momentum operators, $\left[\hat{x},\hat{p}
		\right]=i\hbar$ and the Ehrenfest theorem, the equation of motion for the system is obtained as
		\begin{equation}
		\frac{d^{2}\langle\hat{x}\rangle}{dt^{2}}+\gamma(t)\frac{d\langle\hat{x}\rangle}{dt}+\omega^{2}(t)\langle\hat{x}\rangle=0,
		\label{forceEq}
		\end{equation}
		where, $\gamma(t)=-\frac{d}{dt}\left(ln~f(t)\right)$ and $\langle\hat{x}\rangle$ represents expectation value  of $\hat{x}$. The rate of variation of motion of the above oscillator is determined by the time-varying function $\gamma(t)$, and it is termed as Time Dependent Coefficient of Friction (TDCF). We can rewrite the Hamiltonian in a more familiar and convenient form,
		\begin{equation}
		\mathcal{\hat{H}}=e^{-\Gamma(t)}\frac{\hat{p}^{2}}{2}+e^{\Gamma(t)}\frac{\omega^{2}(t)\hat{x}^{2}}{2},
		\label{ParaHamil}
		\end{equation}
		with $\gamma(t)=\dot{\Gamma}(t)$. The above Hamiltonian implies that the variation in the TDCF of the Oscillator with time will alter both the potential and kinetic energy of the system. In this context, TDCF is the only physically relevant quantity related to a time-varying frictional force in terms of the dimensionless quantity $\Gamma(t)$ (time-varying number), which helps to analyze the system dynamics. TDCF brings the effect of friction into the Hamiltonian. We are not bothered about the bath's exact nature, but the TDCF helps to understand the dissipation due to friction using the above Hamiltonian without incorporating the explicit bath Hamiltonian. We can also observe that the well known dissipative Hamiltonian of Caldirola-Kanai type can be sorted out from the above equation for a constant TDCF~\cite{Caldirola1941,Kanai1948}. Usually, the systems of explicitly time-varying energy due to interaction with the environment are treated as open quantum systems, where the environment's influence is specified by incorporating the bath and interaction Hamiltonian with the system Hamiltonian.  However, in the above formulation, the time-dependent variation of the system's total energy depends on the form of TDCF and the time-dependent frequency without the explicit consideration of the bath (environment) causing it. In other words, the above Hamiltonian is not explicitly considering the bath or interaction Hamiltonian as in the case of open quantum systems, but the influence of the bath is consolidated to the function $\Gamma(t)$.  The Hamiltonian completely describes the system's time evolution with the variation of energy depending on the TDCF. This property of Hamiltonian allows us to follow the invariant based inverse engineering method for STA~\cite{Torrontegui,Guery2019,del2019}.

		The  Lewis-Riesenfeld method of invariants allows to cook up the invariant $\mathcal{\hat{I}}$ for any arbitrary Hamiltonian $\mathcal{\hat{H}}$, by imposing the condition of invariance using the formula~\cite{Lewis1969}
		\begin{equation}
		\frac{\partial \mathcal{\hat{I}}}{\partial t}+\frac{1}{i}\left[\mathcal{\hat{I}},\mathcal{\hat{H}}\right]=0.
		\label{invarianceeq}
		\end{equation}
		Applying this to the Hamiltonian $\mathcal{\hat{H}}$ in equation (\ref{ParaHamil}), we obtain the corresponding Invariant $\mathcal{\hat{I}}$ of the form,
		\begin{equation}
		\mathcal{\hat{I}}=a(t)\hat{x}^{2}+b(t)\left[\hat{x},\hat{p}\right]_{+}+c(t)\hat{p}^{2},	
		\end{equation}
		where, a,b and c are functions of time and $\left[\hat{x},\hat{p}\right]_{+}$ is the anti-commutator of position and momentum operators. Solving for the time-dependent coefficients by using equation (\ref{invarianceeq}) we will be able to deduce the invariant explicitly as (hereafter we represent $\Gamma(t)$ simply as $\Gamma$)
		\begin{equation}
		\mathcal{\hat{I}}=\frac{1}{2}\left\{\left(\frac{x}{\rho}e^{\frac{\Gamma }{2}}\right)^{2}\omega_{0}^{2}+\left(\rho p e^{\frac{-\Gamma }{2}}-\left\{\dot{\rho}-\frac{\dot{\Gamma}\rho}{2}\right\}e^{\frac{\Gamma}{2}}x\right)^{2}\right\}
		\label{invariant}
		\end{equation}
		with time-dependent functions,
		$$a(t)= 2\left\{\left(\dot{\rho}-\frac{\dot{\Gamma}\rho}{2}\right)^{2}+\frac{\omega_{0}^{2}}{\rho^{2}}\right\}e^{\Gamma},$$
		$$b(t)=-2\left\{\rho\dot{\rho}-\frac{\rho^{2}\dot{\Gamma}}{2}\right\},$$
		$$c(t)=2\rho^{2}e^{-\Gamma}.$$
		It can be observed that the formulated $\mathcal{\hat{I}}$ is in the same form as in the ref~\cite{Khandekar1979}, except the presence of a time-dependent perturbative force. The variable  $\rho$ is a function of time and generally called the scaling factor, first introduced by Lewis and  Riesenfeld to scale the invariant equation and later used to control STA using the inverse engineering approach. The necessary condition to be satisfied by the $\rho$ is~\cite{Lewis1969}
		\begin{equation}
		\ddot{\rho}+\Omega^{2}\rho=\frac{\omega_{0}^{2}}{\rho^{3}},
		\label{ermokov}
		\end{equation}
		such that the $\mathcal{\hat{I}}$ will obey equation (\ref{invarianceeq}) with $\mathcal{\hat{H}}$. The above equation is in the form of  Ermakov equations~\cite{John1981}. The so-called shifted frequency $\Omega=\sqrt{\omega^{2}-\frac{\dot{\Gamma}^{2}}{4}-\frac{\ddot{\Gamma}}{2}}$ is influenced by both the TDCF of the oscillation and it's first derivative. The same shift in the frequency is obtained in ref~\cite{Pedrosa1987} using canonical transformation. In the following sections, we will utilize $\mathcal{\hat{I}}$ to implement STA protocols. 
		
	\end{section}
	
	\begin{section}{STA Protocol}
		Shortcut protocols corresponding to the Hamiltonian of equation (\ref{ParaHamil}) for some frequency modulation obeying equation (\ref{ermokov}) can be obtained by using appropriate boundary conditions, which generate the same initial and final states of equilibrium adiabatic process. The boundary conditions are critical in STA, and it decides the form of the scaling factor that could be considered for inverse engineering. The initial and final state of the system under observation is specified through boundary conditions, and it will construct the initial and final structure of the invariant. We have no control over the TDCF, which is a consequence of the interaction with some bath. The inverse engineering can be done by finding the expression for frequency $\omega(t)$ from the  Ermakov equation (\ref{ermokov}) using $\rho(t)$, TDCF ($\dot{\Gamma}$) and its time derivative $\ddot{\Gamma}$. We rewrite the Hamiltonian $\mathcal{\hat{H}}$ (Eq. \ref{ParaHamil}) with the corresponding expression of $\omega(t)$ obtained from Ermakov equation and represent it as $\mathcal{\hat{H}^{IE}}$ where $\mathcal{IE}$ represents `Inverse Engineered'. In this section, we start with the relation between the general solution of time-dependent Schrodinger equation for $\mathcal{\hat{H}}$ and eigenstates of  $\mathcal{\hat{I}}$, then we define the boundary conditions to implement STA and finally obtain the expectation value of energy as a function of time for the shortcut process. 
		
		In general the solution of time-dependent Schrodinger equation of the Hamiltonian $\mathcal{\hat{H}}$ is related to the instantaneous eigenstates of LR Invariant $\mathcal{\hat{I}}$ by a time-dependent phase $\alpha(t)$ as~\cite{Lohe2009}
			\begin{equation}
			\vert\Psi_{n}(t)\rangle=e^{i\alpha(t)}\vert\phi_{n}(t)\rangle,
			\end{equation}
			where, $\vert\Psi_{n}(t)\rangle$ are the general solution of time-dependent Schrodinger equation of Hamiltonian $\mathcal{\hat{H}}$ and  $\vert\phi_{n}(t)\rangle$ are the ortho-normal eigenstates of $\mathcal{\hat{I}}$. As a necessary condition for STA, the set of eigenstates shared by both $\mathcal{\hat{H}}$ and $\mathcal{\hat{I}}$ at initial and final instant of time can be obtained using the commutation relation~\cite{Sakurai1995},
			\begin{equation}
			\left[\mathcal{\hat{H}}\left(0,\tau\right),\mathcal{\hat{I}}\left(0,\tau\right)\right]=0.
			\label{commBC}
			\end{equation} 
			The system should be in the corresponding eigenstate of $\mathcal{\hat{I}}$, before and after the inverse engineered process which satisfies the equation (\ref{commBC}) and make sure the adiabatic transition within a finite time scale~\cite{Chen2010}. Using the expressions for $\mathcal{\hat{H}}$ and $\mathcal{\hat{I}}$ (equations \ref{ParaHamil} $\&$ \ref{invariant}) we can solve for the commutation relation in equation (\ref{commBC}) resulting,
		\begin{equation}
		\left(\dot{\rho}-\frac{\dot{\Gamma}\rho}{2}\right)^{2}+\frac{\omega_{0}^{2}}{\rho^{2}}-\omega^{2}\rho^{2}=0,
		\end{equation}
		\begin{equation}
		\rho\dot{\rho}-\frac{\dot{\Gamma}\rho^{2}}{2}=0.
		\end{equation}
		Solution for the above equations for the initial and final instants of time can be found as
		\begin{equation}
		\rho\left(0\right)=1,
		\label{inbound}
		\end{equation} 
		\begin{equation}
		\rho\left(\tau\right)=\sqrt{\frac{\omega_{0}}{\omega_{\tau}}},
		\label{finalbound}
		\end{equation}
		\begin{equation}
		\dot{\rho}(0)=\frac{\dot{\Gamma}(0)}{2}
		\label{rhodotzero}
		\end{equation}
		and
		\begin{equation}
		\dot{\rho}(\tau)=\frac{\dot{\Gamma}(\tau)\sqrt{\frac{\omega_{0}}{\omega_{\tau}}}}{2},
		\label{rhodottau}
		\end{equation}
		where, $0$ is the initial time, and $\tau$ is the final time. The conditions (\ref{inbound}) and (\ref{finalbound}) are the same for the STA of QHO~\cite{Chen2010,Obinna2018,Obinna2019} and defined with the initial and final values of time-dependent frequency of the oscillator. However, to fix the values of the time derivative of scaling factor at $t=0$ and $t=\tau$ (\ref{rhodotzero},\ref{rhodottau}), the initial and final values of the TDCF is necessary, which shows the effect of time-dependent friction on the shortcut process. 
		
		In an adiabatic process, the system is isolated, and any change in the system's energy levels is considered work done by or on the system. We can consider the Harmonic oscillator under time-dependent frictional force as an isolated entity during the adiabatic process, and the STA process is achieved by the evolution of such an isolated entity under the inverse engineered Hamiltonian $\mathcal{\hat{H}^{IE}}$ with modified frequency as resulting from equation (\ref{ermokov}). The expectation value $\langle\mathcal{\hat{H}^{IE}}\rangle$ in the STA path is obtained by operating the eigenstates of  $\mathcal{\hat{I}}$ with $\mathcal{\hat{H}^{IE}}$. We can simplify the above mathematical process by rewriting the Hamiltonian $\mathcal{\hat{H}}$ in equation (\ref{ParaHamil}) as $\mathcal{\hat{H}^{IE}}$ in terms of 
		\begin{equation}
		\hat{X}=\frac{\sqrt{\omega_{0}}e^{\frac{\Gamma}{2}}}{\rho}\hat{x}
		\end{equation}
		and 
		\begin{equation}
		\hat{P}=\frac{\rho e^{\frac{-\Gamma}{2}}}{\sqrt{\omega_{0}}}\hat{p}+\frac{\left(\dot{\rho}-\frac{\dot{\Gamma}\rho}{2}\right) e^{\frac{\Gamma}{2}}}{\sqrt{\omega_{0}}}\hat{x},
		\end{equation}
		where $\hat{X}$ and $\hat{P}$ are the position and momentum operators of the invariant $\mathcal{\hat{I}}$ in equation~(\ref{invariant}) and $\omega_{0}$ is the initial frequency of the QHO under time-dependent friction. The form of $\mathcal{\hat{H}^{IE}}$ obtained on the above substitution is
			\begin{equation}
			\mathcal{\hat{H}^{IE}}=\frac{1}{2}\left(\frac{\omega_{0}}{\rho^{2}}\hat{P}\hat{P}^{\dagger}+\frac{\left(\dot{\rho}-\frac{\dot{\Gamma}\rho}{2}\right)^{2}+\omega^{2}\rho^{2}}{\omega_{0}}\hat{X}^{2}-2\frac{\left(\dot{\rho}-\frac{\dot{\Gamma}\rho}{2}\right)}{\rho}\hat{P}\hat{X} \right).	
			\label{InverseEngineeredHamiltonian}
			\end{equation}
		Then, by using the creation and annihilation operators $\hat{a}^{\dagger}$ and $\hat{a}$ defined as 
		\begin{equation}
		\hat{a}^{\dagger}=\frac{1}{\sqrt{2}}\left(\hat{X}-i\hat{P}\right)
		\label{creation}
		\end{equation}
		\begin{equation}
		\hat{a}=\frac{1}{\sqrt{2}}\left(\hat{X}+i\hat{P}\right),
		\label{annihil}
		\end{equation}
		with properties
		$$\hat{a}^{\dagger}\vert \phi_{n}(t)\rangle=\sqrt{n+1}\vert\phi_{n+1}(t) \rangle$$
		$$\hat{a}\vert \phi_{n}(t)\rangle=\sqrt{n}\vert \phi_{n-1}(t)\rangle$$
		on the eigenstates $\vert \phi_{n}(t)\rangle$ of the invariant $\mathcal{\hat{I}}$, the expectation value $\langle\mathcal{\hat{H}^{IE}}\rangle$ is obtained by combining equations (\ref{InverseEngineeredHamiltonian}), (\ref{creation}) and (\ref{annihil}) as
		\begin{equation}
		\langle\mathcal{\hat{H}^{IE}}\rangle=\langle\phi_{n}(t)\vert\mathcal{\hat{H}^{IE}}\vert\phi_{n}(t)\rangle=\frac{\left(2n+1\right)}{4\omega_{0}}\left(\frac{\omega_{0}^{2}}{\rho^{2}}+\left[\dot{\rho}-\frac{\dot{\Gamma}\rho}{2}\right]^{2}+\omega^{2}\rho^{2}\right),
		\label{InvEngExpecnVal}
		\end{equation}
		where we have used the properties of annihilation and creation operators and ortho-normal property of $\vert\phi_{n}(t)\rangle$, which simplifies the problem without the use of the explicit form of eigenstates of the Invariant. The invariant in equation (\ref{invariant}) along with the boundary conditions in equations (\ref{inbound}) to (\ref{rhodottau}) can be used to drive the QHO under time-dependent frictional force for any arbitrary variation in the TDCF. This variation in TDCF can be formulated as a function from the experimental results of energy variation during the adiabatic time evolution of any system governed by the Hamiltonian $\mathcal{\hat{H}}$ in equation (\ref{ParaHamil}). There is another possibility of attaining the adiabatic states of ideal QHO from STA driving of the QHO with time-dependent friction. The above possibility is achieved by modifying the boundary conditions for the scaling factor so that the inverse engineered Hamiltonian in equation (\ref{InverseEngineeredHamiltonian}) results in the adiabatic states of the ideal QHO. The above strategy depends on the form of TDCF as it is a determining factor of the boundary conditions (Equation (\ref{rhodotzero}) and (\ref{rhodottau})). In the following section, we investigate the systems with a particular class of TDCF, for which we can drive the system from one initial quantum state of ideal time-dependent QHO to the corresponding adiabatic final state in a finite time interval. 				
		
	\end{section}
	
	\begin{section}{STA for Harmonic Oscillator}
		The commutation relation in equation (\ref{commBC}) shows that the achieved state through STA corresponds to the QHO under time-dependent friction. However, the expectation value of the inverse engineered Hamiltonian (\ref{InvEngExpecnVal}) shows a similar form of the energy eigenvalues for the ideal QHO at initial and final instant of time. We seek for a particular class of QHO under time-dependent friction, (a particular form of TDCF) for which the use of the invariant $\mathcal{\hat{I}}$ (\ref{invariant}) along with boundary conditions (\ref{inbound}-\ref{rhodottau}) defined for scaling factor, drive the system from an initial state to the corresponding adiabatic final state of an ideal QHO. From the Hamiltonian $\mathcal{\hat{H}}$ (\ref{ParaHamil}), it is evident that the absence of the time-dependent friction at initial and final instant of time gives back the ideal QHO Hamiltonian at the starting and the ending of the process, irrespective of the presence of TDCF at intermediate times. Also, the commutation relation (\ref{commBC}) gives the boundary conditions for ideal QHO. This particular situation of friction at intermediate times except at the initial and final instants of time is given by the boundary condition for $\Gamma(t)$ as
			\begin{figure}[tb]
				\centering
				\includegraphics[width=.6\linewidth]{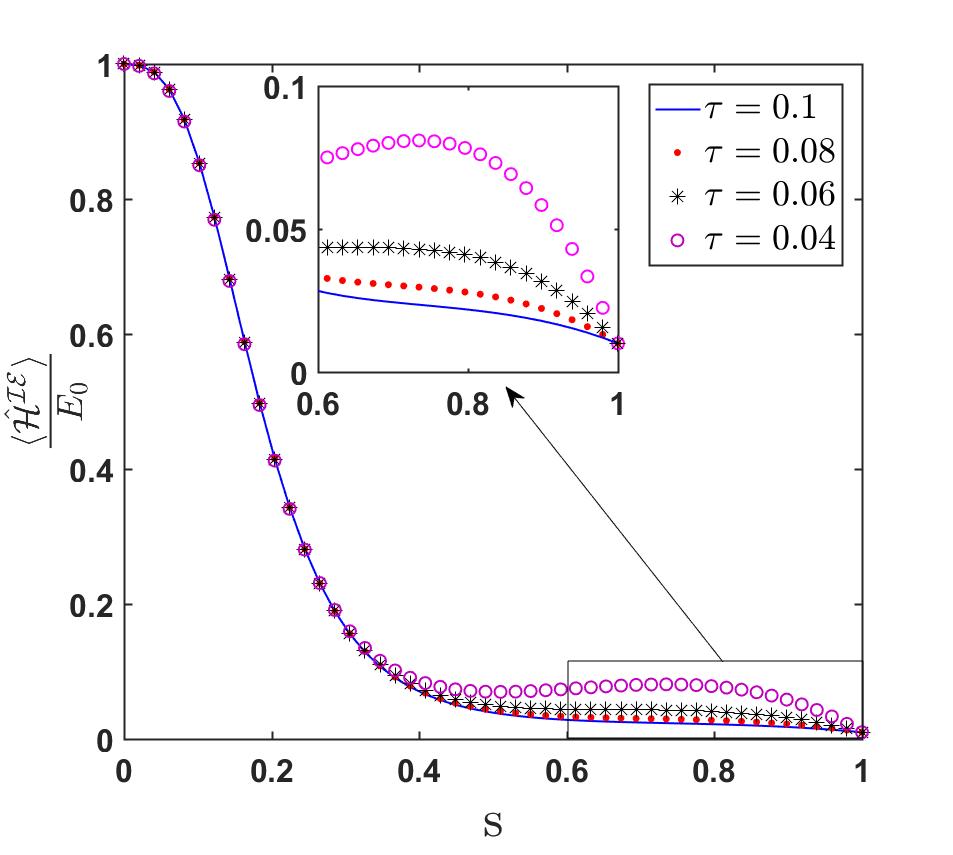}
				\caption{Variation of $\langle\hat{\mathcal{H}}^{\mathcal{IE}}\rangle$ is plotted as a dimension less fraction with initial energy of Harmonic Oscillator $E_{0}$ to $s=\frac{t}{\tau}$ for different values of $\tau$ and $\omega_{0}=250\times2\pi$, $\omega_{\tau}=2.5\times2\pi$, $n=0$.}
				\label{ExpHIE}
			\end{figure}
			\begin{equation}
			\Gamma(0,\tau)=0,
			\label{boundaryGamma}
			\end{equation}
			which modifies the commutation relation in equation (\ref{commBC}) and gives,
			\begin{equation}
			\left[\mathcal{\hat{H}}(0,\tau),\mathcal{\hat{I}}(0,\tau)\right]=\left[\mathcal{\hat{H}}_{qho}(0,\tau),\mathcal{\hat{I}}(0,\tau)\right]=0,
			\label{commrel2}
			\end{equation}
			where, $\mathcal{\hat{H}}_{qho}=\frac{\hat{p}^{2}}{2}+\frac{\omega^{2}(t)\hat{x}^{2}}{2}$ is the Hamiltonian for QHO.
		The equation (\ref{commrel2}) indicate that the STA process using the invariant $\mathcal{\hat{I}}$ with all the previously defined boundary conditions for scaling factor drives a system defined by the Hamiltonian $\mathcal{\hat{H}}$ with some arbitrary time-dependent function $\Gamma(t)$ obeying the boundary condition in equation (\ref{boundaryGamma}) for required initial and final adiabatic states of QHO. The consequence of the boundary condition (\ref{boundaryGamma}) is that the Hamiltonian $\mathcal{\hat{H}}$ takes the form of QHO Hamiltonian at both the ends of the process. In the above scenario, the expectation value of the inverse engineered Hamiltonian should converge to the energy of QHO at the beginning and ending of the STA process, and it is evident from the equation (\ref{InvEngExpecnVal}) that
		\begin{equation}
			\langle\mathcal{\hat{H}^{IE}}(0,\tau)\rangle=\langle\mathcal{\hat{H}}_{qho}(0,\tau)\rangle=\left(n+\frac{1}{2}\right)\omega(0,\tau).
			\label{ExpHarOsc}
		\end{equation} 
		We can in general set up a STA protocol for QHO from this class of Hamiltonian $\mathcal{\hat{H}}$ with TDCF ($\dot{\Gamma}(t)$) obeying the condition $\int \dot{\Gamma}(t)dt=0$ at initial and final instant of time. This method of obtaining STA for QHO applies to any arbitrary physical interaction of the system with the bath obeying the condition (\ref{boundaryGamma}). However, for numerical analysis and verification of the above-derived method, we assume a Hamiltonian $\mathcal{\hat{H}}$ with a particular TDCF corresponding to the numerical function
			\begin{equation}
			\Gamma(t)=s^{3}(s-1)^{3},
			\label{gammafun}
			\end{equation}
			where, $s=\frac{t}{\tau}$ makes the function dimensionless. The above function obeys the boundary condition in equation (\ref{boundaryGamma}) and also gives the boundary conditions for the scaling factor from equations (\ref{inbound}-\ref{rhodottau}) as
			$$\rho(0)=1,\rho(\tau)=\sqrt{\frac{\omega_{0}}{\omega_{\tau}}}$$
			$$ \dot{\rho}(0)=0, \dot{\rho}(\tau)=0.$$  
			According to the above boundary conditions, we choose a well known scaling factor~\cite{Chen2010} 
		\begin{equation}
		\rho(t)=6 \left( \sqrt{\frac{\omega_{0}}{\omega_{\tau}}}-1\right) s^{5}-15\left(\sqrt{\frac{\omega_{0}}{\omega_{\tau}}}-1\right) s^{4}+10\left(\sqrt{\frac{\omega_{0}}{\omega_{\tau}}}-1\right) s^{3}+1.
		\label{scalefac}
		\end{equation}
		This specific function is used in most applications ( Quantum Otto engines, Atomic transport, etc.) of STA protocols to drive the QHO using the invariant method. Using the above scaling factor, we can inverse engineer the Hamiltonian $\mathcal{\hat{H}}$ and proceed with the method explained in previous sections to achieve STA for QHO. Numerical analysis can be made for various parameters of the STA process, including the cost of implementation.
		
		Numerical computation of expectation values of energy in the STA process, for various final times, can be done using equation (\ref{InvEngExpecnVal}). Figure \ref{ExpHIE} shows the variation of the expectation value of Inverse engineered Hamiltonian in the above setting for STA of Harmonic Oscillator. We have plotted the dimensionless ratio of the expectation value of $\mathcal{\hat{H}^{IE}}$ to initial energy of the Harmonic Oscillator against $s=\frac{t}{\tau}$ for various final times $\tau$. It is assumed that the system was thermalized to the ground state before the shortcut process, where the system expands within a short time $\tau$. During this expansion process, the oscillator frequency will change from a higher value to a lower value. For numerical calculation, we have selected the frequency change from $250\times2\pi$ Hz to $2.5\times2\pi$ Hz, which is experimentally executable and less sensitive to random noise~\cite{Chen2010}. It is theoretically possible to reach a short time durations till the trap inversion. However, the chosen range of frequency for the fewer noise effects causes a very high cost for shorter durations of time. We can achieve much shorter durations of time for high frequencies but with even more cost.
		
		\begin{figure}[tb]
			\centering
			\begin{subfigure}{.45\linewidth}
				\includegraphics[scale=.2]{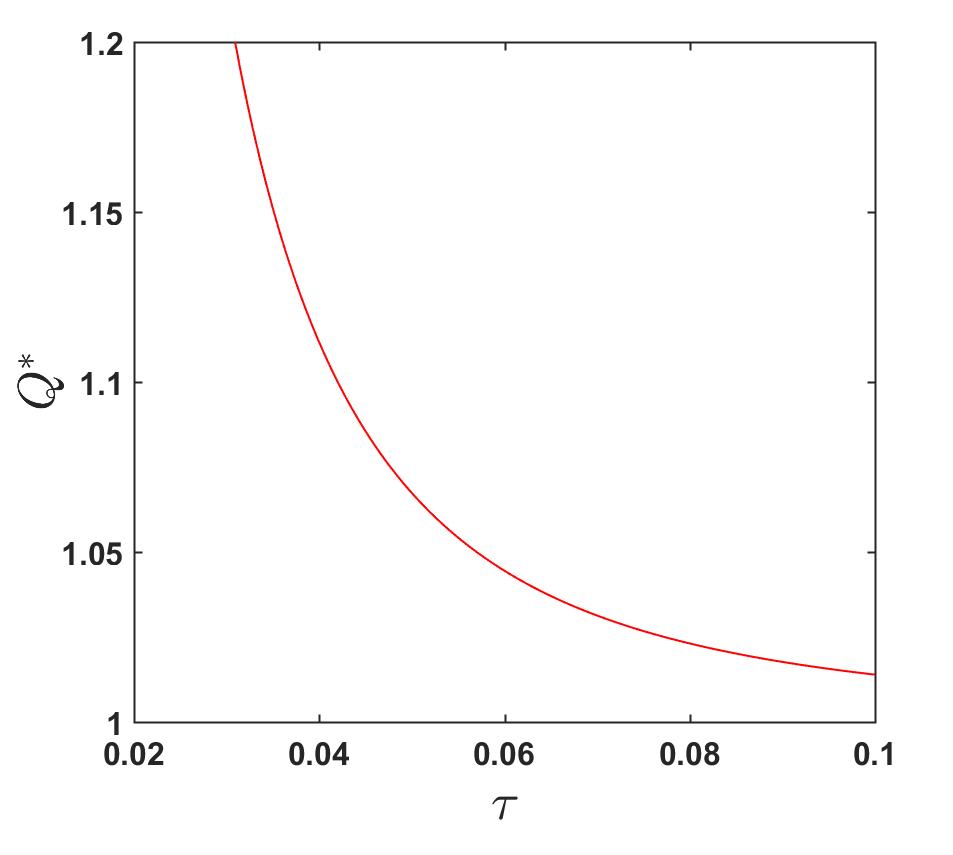}
				\caption{}
				\label{QAvevalue}
			\end{subfigure}
			\begin{subfigure}{.45\linewidth}
				\includegraphics[scale=.2]{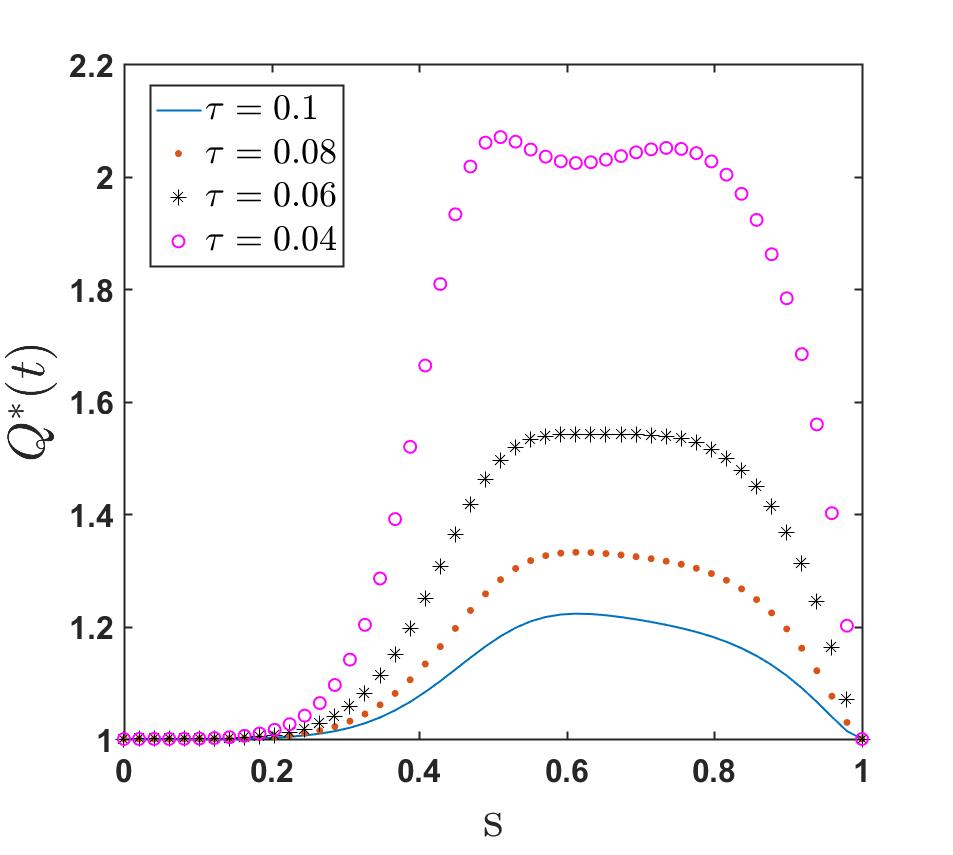}
				\caption{}
				\label{AdiaPar}
			\end{subfigure}
			\begin{subfigure}{.45\linewidth}
				\includegraphics[scale=.2]{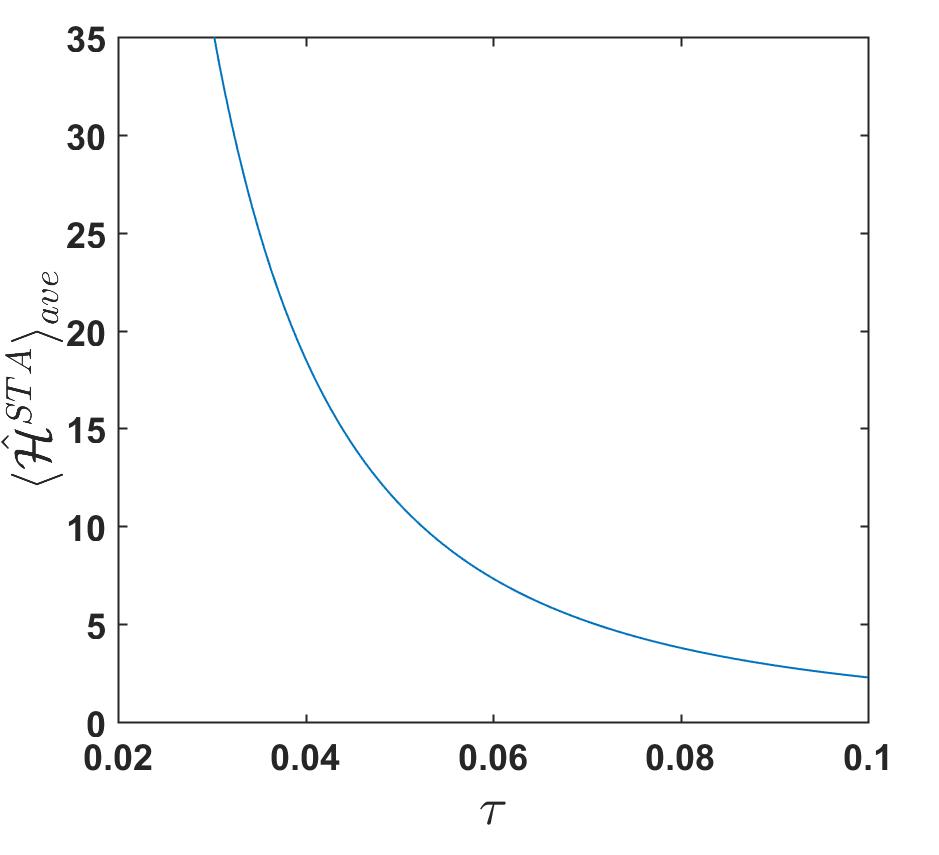}
				\caption{}
				\label{HstaAvevalue}
			\end{subfigure}
			\begin{subfigure}{.45\linewidth}
				\includegraphics[scale=.2]{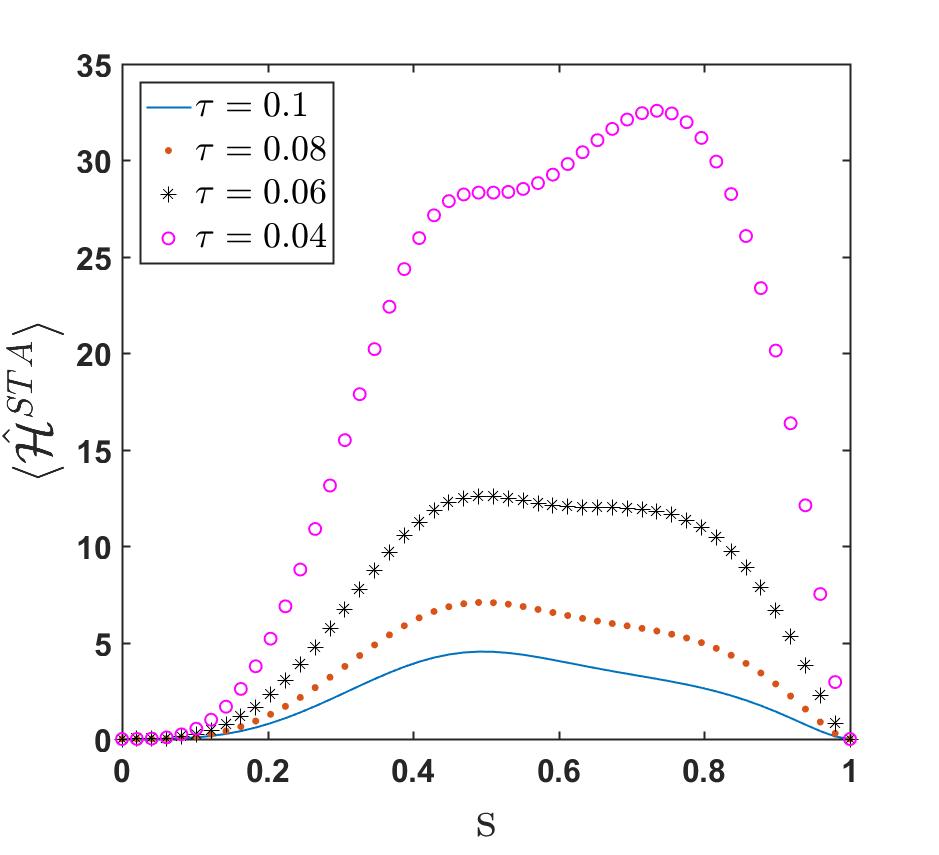}
				\caption{}
				\label{Hstavalue}
			\end{subfigure}
			\caption{Plots for (a) Variation of adiabaticity parameter $Q^{*}$ for various final times $\tau$, (b) instantaneous behaviour of Adiabaticity parameter $Q^{*}(t)$ for various final times $\tau$ is plotted against $s=\frac{t}{\tau}$, (c) average energy cost $\langle\hat{H}^{STA}\rangle_{ave}$ for various final times $\tau$ and (d) instantaneous energy cost $\langle\hat{H}^{STA}\rangle$ for various final times $\tau$ is plotted against $s=\frac{t}{\tau}$.}
		\end{figure}

		Adiabaticity parameter $Q^{*}$ is a measure of adiabaticity of the shortcut process and defined as the ratio of average energy of the shortcut process $\langle\hat{\mathcal{H}}^{\mathcal{IE}}\rangle_{ave}$ to the average adiabatic energy $\langle\hat{\mathcal{H}}_{qho}\rangle_{ave}$~\cite{Deffner2010,Kdi1953}
		\begin{equation}
		Q^{*}=\frac{\langle\hat{\mathcal{H}}^{\mathcal{IE}}\rangle_{ave}}{\langle\hat{\mathcal{H}}_{qho}\rangle_{ave}}=\frac{\int_{0}^{\tau}\langle\hat{\mathcal{H}}^{\mathcal{IE}}(t^{\prime})\rangle  d t^{\prime}}{\int_{0}^{\tau}\langle\hat{\mathcal{H}}_{qho}(t^{\prime})\rangle d t^{\prime}},
		\end{equation}
		where $\langle~~\rangle_{ave}$ represents the time average of expectation values of corresponding Hamiltonian. The instantaneous behavior of the Adiabaticity parameter can be analyzed numerically using 
		$$Q^{*}(t)=\frac{\langle\hat{H}^{IE}\rangle}{\langle\hat{H}_{qho}\rangle}$$
		\begin{equation}
		=\frac{1}{2\omega_{0}\omega}\left(\frac{\omega_{0}^{2}}{\rho^{2}}+\left(\dot{\rho}-\frac{\dot{\Gamma}\rho}{2}\right)^{2}+\omega^{2}\rho^{2}\right).
		\label{Instadiapar}
		\end{equation}
		Adiabaticity parameter is plotted in Figure \ref{QAvevalue} and its instantaneous behavior is plotted in Figure \ref{AdiaPar} for all the other variables specified in Figure \ref{ExpHIE}. It is evident from the plot \ref{QAvevalue} that the adiabaticity parameter tends to 1 for large times scales; thus, the process tends to be completely adiabatic as $\tau$ increases. Instantaneous behavior of the Adiabaticity parameter varies from the initial value 1 to the final value 1 to ensure the adiabatic final states and the value deviate from the adiabatic trajectory at intermediate times. This deviation can be negligible by appropriate control of the dynamics with proper designing of the scaling factor.
		
		The whole shortcut process is done by modifying the Hamiltonian; thus, it can bring back the adiabatic final states within a short time. There must be a cost for such deviation in the dynamics of the process from the actual adiabatic dynamics. This cost can be measured as the difference between the expectation value of energy for inverse engineered Hamiltonian and that of harmonic oscillator Hamiltonian at any instant of time, and it is given by the formula~\cite{Obinna2018},
		
		\begin{equation}
		\langle\mathcal{\hat{H}}^{STA}\rangle=\langle\mathcal{\hat{H}^{IE}}\rangle-\langle\mathcal{\hat{H}}_{qho}\rangle=\frac{\left(2n+1\right)}{4\omega_{0}}\left(\frac{\omega_{0}^{2}}{\rho^{2}}+\left(\dot{\rho}-\frac{\dot{\Gamma}\rho}{2}\right)^{2}+\omega^{2}\rho^{2}-2\omega_{0}\omega\right),
		\end{equation}
		which is plotted in Figure \ref{Hstavalue}. This implementation cost is very high for small time scales of the shortcut, making it difficult to achieve very short processes. As the shortcut process's initial and final energy coincides with that of the actual adiabatic process, the cost value is zero for the process's endpoints. The average value of implementation cost can be found for any final time $\tau$ by time-averaging the expectation value as $$\langle\mathcal{\hat{H}}^{STA}\rangle_{ave}=\left(\frac{1}{\tau}\right)\int_{0}^{\tau}\langle\mathcal{\hat{H}}^{STA}\rangle dt$$
		and it is plotted in Figure \ref{HstaAvevalue}, which shows a gradual decrease in implementation cost and it tends to zero for long time scales implying that the process is equivalent to the equilibrium adiabatic process without any control over the system for large $\tau$.
		\begin{figure}[tb]
			\centering
			\includegraphics[width=.9\linewidth]{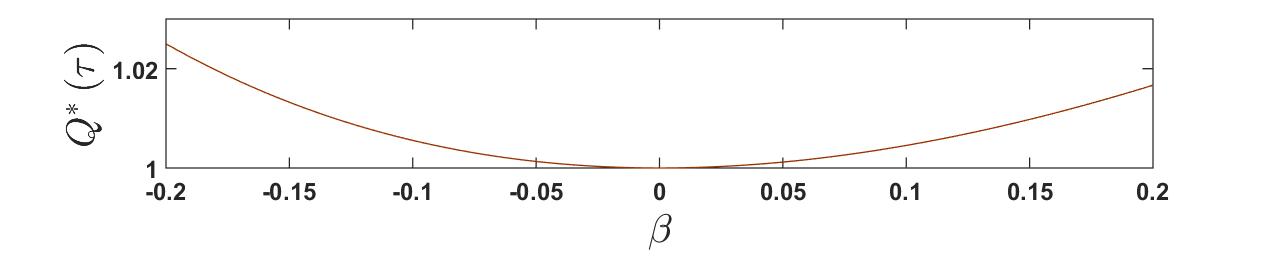}
			\caption{Variation of $Q^{*}(\tau)$ is plotted for different values of $\beta$}
			\label{InErrorCalc}
		\end{figure}
		The error-free implementation of the STA protocol is essential for the expected outcome. There are several studies dedicated to the optimization of the STA protocol with high stability against various systematic and random errors~\cite{Guery2019}. We selected the frequency range, which is less sensitive to the random noise, but we cannot ignore the chances of systematic errors. The inverse engineered Hamiltonian in the presence of the systematic error is 
		\begin{equation}
			\mathcal{\hat{H}^{IE}}_{s}=\mathcal{\hat{H}^{IE}}+\beta \mathcal{\hat{H}}_{p},
		\end{equation}
		where $\mathcal{\hat{H}}_{p}$ is the perturbation part due to error and $\beta$ is the amplitude of the systematic error~\cite{Rusch2012}. We check the stability of our protocol by constraining to the perturbation that causes a shift in the frequency $\omega$ (obtained from the Ermakov equation~(\ref{ermokov})), without affecting the form of $\mathcal{\hat{H}^{IE}}$. Also, assuming the shift in frequency is of the form $\left(1+\beta\right)\omega$, which induces changes in the adiabaticity parameter (Equation (\ref{Instadiapar})). The analysis of variation of $Q^{*}(\tau)$ from unity at the final instant of time for various amplitudes of systematic error $\beta$ is shown in Figure \ref{InErrorCalc}.  
		The comparison of the results with the definition of adiabaticity parameter shows around 2\% of deviation in the expectation value of inverse engineered Hamiltonian from the required adiabatic value, for a 20\% shift in the frequency of the inverse engineered Hamiltonian, which is highly stable. According to the above result, a well-calibrated experimental setup might result in a negligible error in the output. Still, we can optimize the protocol by redefining the boundary conditions for zero sensitivity to the systematic errors~\cite{Rusch2012}. We analyze the protocol's stability for different types of errors and optimization of the protocol in future works.
		
	\end{section}
	
	\begin{section}{General Approach to Mass Variation}
		The Hamiltonian $\mathcal{\hat{H}}$ (Eq. \ref{ParaHamil}) considered so far is worth studying as it stands for yet another physically relevant and distinct situation, where the mass of the observed system varies with time. Controlling the dynamics of any system by arbitrarily varying its mass is unrealistic, but controlling the system with inherent mass variation is realistic. The interpretation of mass variation in harmonic oscillator Hamiltonian is also found useful to model the optical lattices, identifying the mass as a function of propagation distance~\cite{Rodr2014,Oztas2016}. A shortcut mechanism for such a model is proposed using invariants for Hamiltonian of forced oscillators with varying mass and frequency~\cite{Stefanatos2014}. The shortcut process to design an optical lattice for desired output explained in Ref.~\cite{Stefanatos2014} deals with the coherent final states using an invariant defined for constant mass and using boundary conditions explicitly declared for the required coherent final state. Below, we discuss the usefulness of our approach (developed for QHO under time-dependent friction) in the case of a harmonic oscillator with varying mass $M(\xi)$ and frequency $\omega(\xi)$ for some parameter $\xi$. Our method's generality helps us define an invariant, which is general for any arbitrary variation in mass and frequency. Also, the boundary conditions defined in analogy with QHO under time-dependent friction can be used to attain any required adiabatic final state in general. The Hamiltonian for the harmonic oscillator with varying mass and frequency can be represented as
		\begin{equation}
		\mathcal{\hat{H}}^{\prime}=\frac{\hat{p}^{2}}{2 M(\xi)}+\frac{M(\xi) \omega^{2}(\xi)\hat{x}^{2}}{2}.
		\label{masshamil}
		\end{equation}
		Comparing Equation (\ref{masshamil}) with (\ref{ParaHamil}) gives~\cite{Pedrosa1997},
		\begin{equation}
		\dot{\Gamma}(\xi)=\frac{d}{d\xi}\left[\ln M(\xi)\right].
		\label{gammaMass}
		\end{equation}
		From equation (\ref{invariant}), the invariant in terms of $M(\xi)$ is 
		\begin{equation}
		\mathcal{\hat{I}}^{\prime}=\frac{1}{2}\left(\left(\frac{x}{\rho^{\prime}}\right)^{2}M(\xi)\omega_{0}^{2}+\frac{1}{M(\xi)}\left(\rho^{\prime} p -\left(M(\xi)\dot{\rho^{\prime}}-\frac{\dot{M}(\xi)\rho^{\prime}}{2}\right)x\right)^{2}\right).
		\label{invariantM}
		\end{equation}
		This invariant is the exact invariant for $\mathcal{H}^{\prime}$, which is different from the one considered in ref~\cite{Stefanatos2014,Pedrosa1997} and the existence of this invariant depends on the Ermakov equation
		
		\begin{equation}
		\ddot{\rho^{\prime}}+\Omega^{\prime 2}\rho^{\prime}=\frac{\omega_{0}^{2}}{\rho^{\prime 3}},
		\label{ermokovM}
		\end{equation}
		where the new shift in the frequency is $\Omega^{\prime}=\sqrt{\omega\left(\xi\right)^{2}+\left(\frac{\dot{M}(\xi)}{2M(\xi)}\right)^{2}-\frac{\ddot{M}(\xi)}{2M(\xi)}}$ and a general STA protocol can be formulated as discussed in section 3. Implementation of the protocol requires the knowledge of the variation of mass with corresponding parameter (time, length, etc.), which decides the particular form of scaling factor $\rho^{\prime}$ for appropriate STA boundary conditions. The general boundary conditions for STA for QHO with $\xi$ dependent mass and frequency, obtained from equations (\ref{inbound}-\ref{rhodottau}) are 
		$$\rho\left(0\right)=1$$
		$$\rho\left(\xi_{\tau}\right)=\sqrt{\frac{\omega\left(0\right)}{\omega\left(\xi_{\tau}\right)}}$$
		$$\dot{\rho}(0)=\left[\frac{d}{d\xi}\left(\ln \sqrt{M(\xi)}\right)\right]_{\xi=0}$$
		$$\dot{\rho}(\xi_{\tau})=\sqrt{\frac{\omega\left(0\right)}{\omega\left(\xi_{\tau}\right)}}\left[\frac{d}{d\xi}\left(\ln \sqrt{M(\xi)}\right)\right]_{\xi=\xi_{\tau}},$$
		where, $\xi_{\tau}$ is the final value of the parameter $\xi$. We used the Ermakov equation (\ref{ermokovM}) to inverse engineer the frequency to drive the system in a shortcut path. The mass variation directly influences the oscillator frequency, which will be evident on the inversion of the Ermakov equation to construct frequency variation $\omega(\xi)$, which is in good agreement with some of the existing interpretations of shortcuts~\cite{Shumpei2008}. On comparison with equation (\ref{InvEngExpecnVal}), the expectation value of energy on the inverse engineered shortcut path is,
		\begin{equation}
		\langle\mathcal{\hat{H}^{\prime}_{ IE}}\rangle=\frac{\left(2n+1\right)}{4\omega_{0}}\left(\frac{\omega_{0}^{2}}{\rho^{2}}+\left(\dot{\rho}-\frac{\dot{M}(\xi)\rho}{2M(\xi)}\right)^{2}+\omega^{2}\rho^{2}\right).
		\label{InvEngExpecnValM}
		\end{equation}	
		\subsection{Photonic Lattice as Mass Varying Hamiltonian}
		The photonic lattice model proposed in ref~\cite{Rodr2014} is semi-infinite and composed of individual waveguides, whose index of refraction varies linearly. It can be modeled as a harmonic oscillator with mass $M(z)$ and frequency $\Omega(z)$, where z is the propagation distance~\cite{Rodr2014,Stefanatos2014}. Considering the field amplitude at n$^{th}$ waveguide as $C_{n}(z)$, $a_{0}(z)$ to modulate linear variation of the refractive index and $a_{1}(z)$,$a_{2}(z)$ as first and second coupling functions. The differential set describes the lattice,
		\begin{equation}
		i\frac{\partial C_{n}(z)}{\partial z}+a_{0}(z)nC_{n}(z)+a_{1}(z)\left[f_{n+1}C_{n+1}(z)+f_{n}C_{n-1}(z)\right]+a_{2}(z)\left[g_{n+2}C_{n+2}(z)+g_{n}C_{n-2}(z)\right]=0,
		\label{latticeEq}
		\end{equation}
		\begin{figure}[tb]
			\centering
			\includegraphics[width=.6\linewidth]{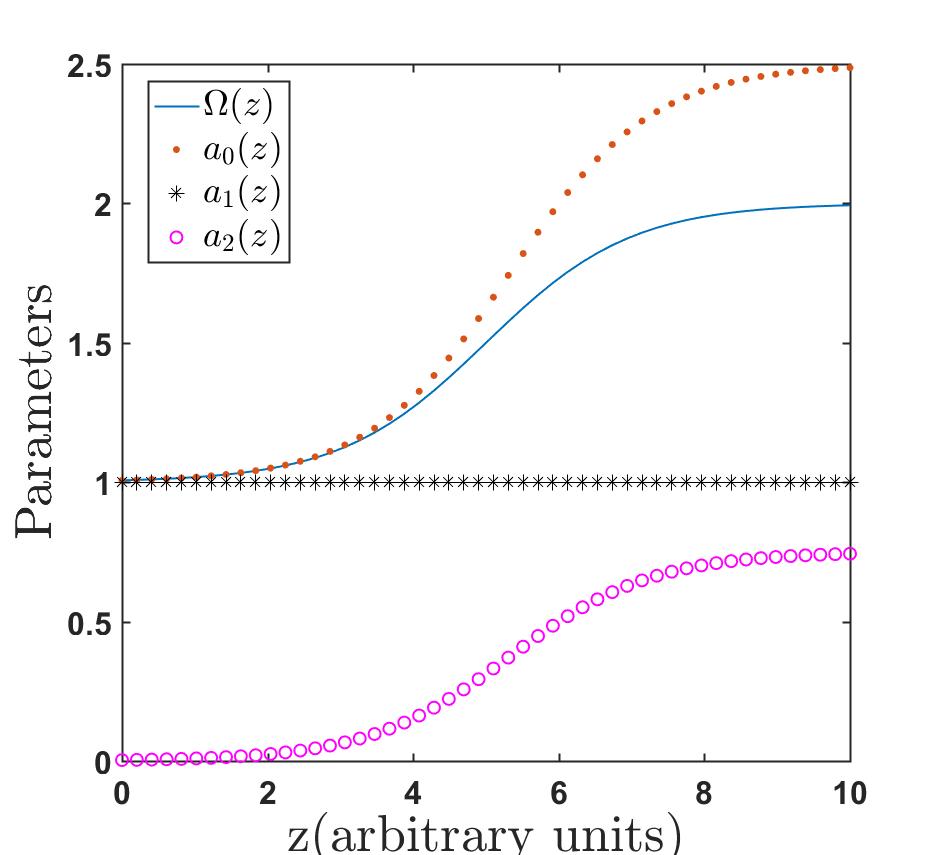}
			\caption{Variation of the parameters $\Omega(z)$, $a_{0}(z)$, $a_{1}(z)$ and $a_{2}(z)$ against the propagation distance $z$ in arbitrary units is similar to one given in ref~\cite{Rodr2014} with $\epsilon=0.5$ and $z_{s}=5$.}
			\label{varparaZ}
		\end{figure}
		\begin{figure}[tb]
			\centering
			\begin{subfigure}{.45\linewidth}
				\includegraphics[scale=.2]{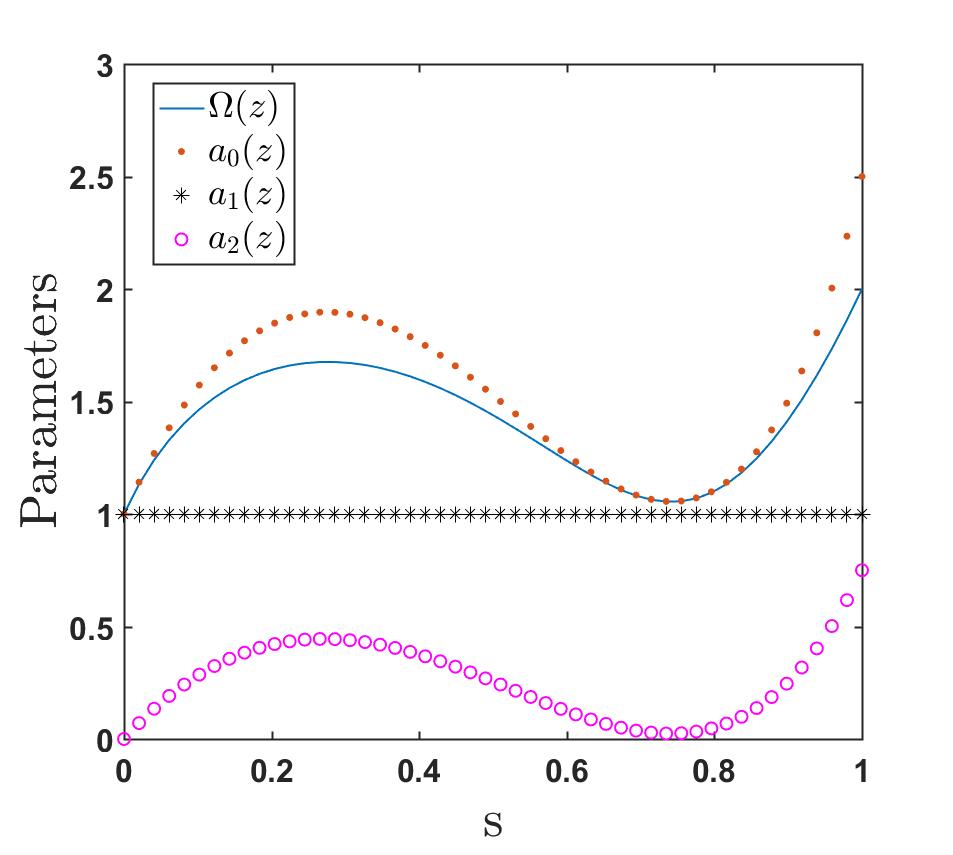}
				\caption{}
				\label{varparasta1}
			\end{subfigure}
			\begin{subfigure}{.45\linewidth}
				\includegraphics[scale=.2]{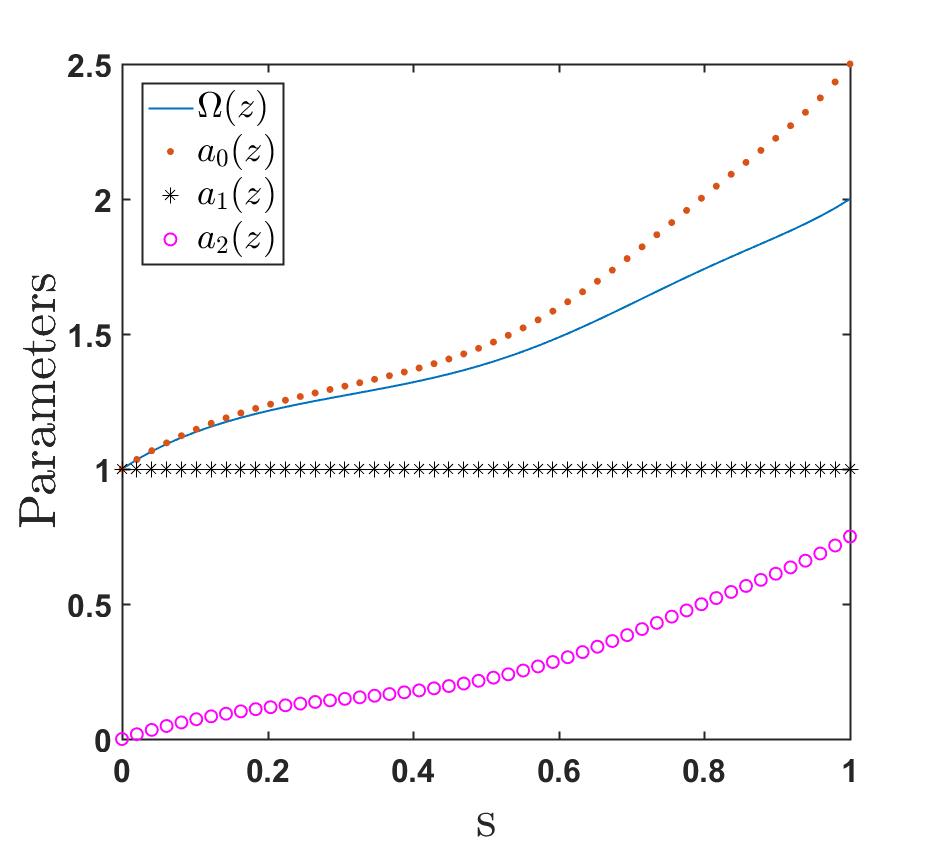}
				\caption{}
				\label{varparasta2}
			\end{subfigure}
			\begin{subfigure}{.45\linewidth}
				\includegraphics[scale=.2]{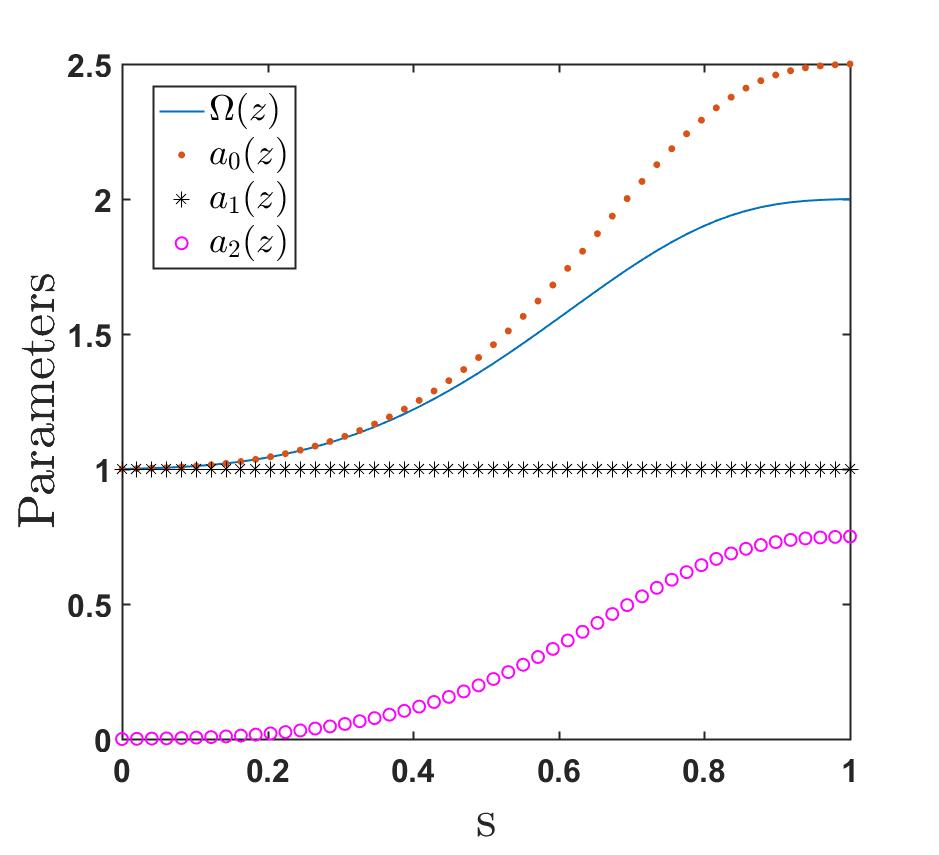}
				\caption{}
				\label{varparasta10}
			\end{subfigure}
			\begin{subfigure}{.45\linewidth}
				\includegraphics[scale=.2]{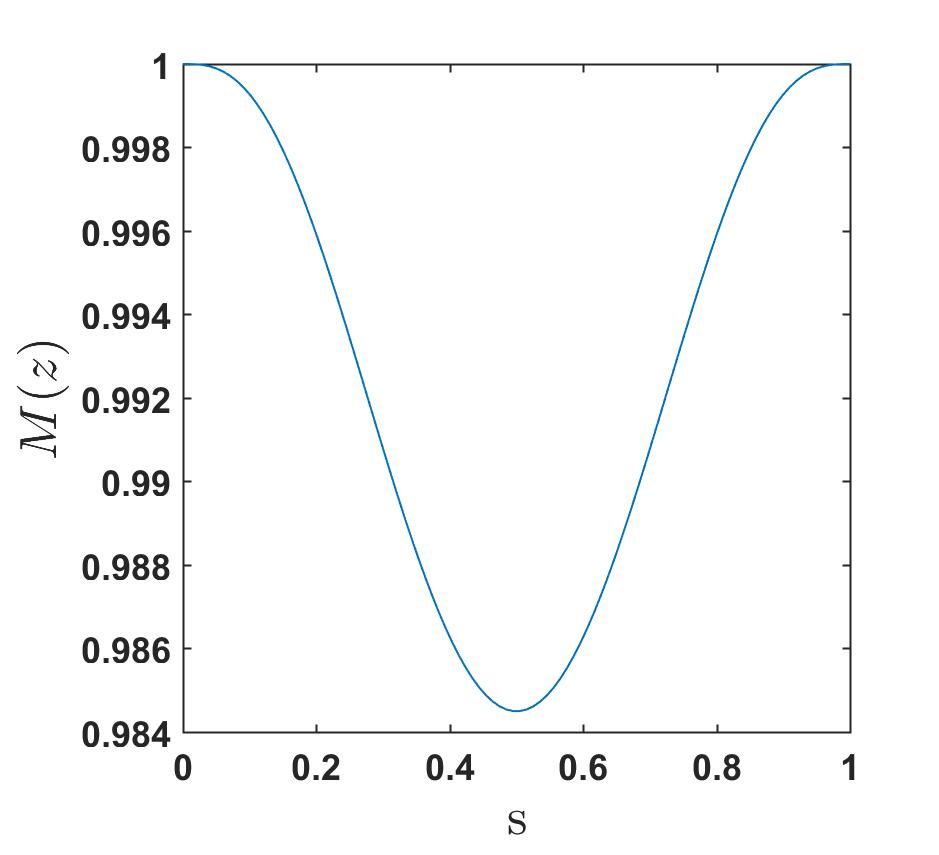}
				\caption{}
				\label{Massvar}
			\end{subfigure}
			\caption{The Variation of parameters $\Omega(z)$, $a_{0}(z)$, $a_{1}(z)$, $a_{2}(z)$ using STA is plotted against $s=\frac{z}{z_{\tau}}$ for the different final propagation distances (a) $z_{\tau}=1$, (b) $z_{\tau}=2$ and (c) $z_{\tau}=10$. The plot (d) shows the mass variation in STA process using equation (\ref{MEq}). We have used the initial and final parameters, $\Omega(0)=M(0)=1$, $\Omega(z_{\tau})=2$ and $M(z_{\tau})=1$ for all the cases.}
		\end{figure}
		where $f_{n}=\sqrt{n}$ and $g_{n}=\sqrt{n(n-1)}$ are the functions of the positions $n=0,1,2...$ of the waveguides in the array and $C_{n}(z)=0$ for $n<0$. If we define a wavefunction $\vert\Psi(z)\rangle=\sum_{j=0}^{n}C_{j}(z)\vert j\rangle$ using the field amplitude $C_{j}$ at j$^{th}$ waveguide, we can rewrite the equation (\ref{latticeEq}) as Schrodinger-like equation~\cite{Rodr2014},
		\begin{equation}
		\mathcal{H}^{\prime}\vert\Psi(z)\rangle=i\frac{\partial \vert\Psi(z)\rangle}{\partial z}
		\label{ShcrLikeEq}
		\end{equation}
		and the corresponding Hamiltonian in terms of annihilation $(\hat{a}\vert n\rangle=\sqrt{n}\vert n-1\rangle)$ and creation $(\hat{a}^{\dagger}\vert n\rangle=\sqrt{n+1}\vert n+1\rangle)$ operators is
		\begin{equation}
		\mathcal{H}^{\prime}=-\left[a_{0}(z)\hat{a}\hat{a}^{\dagger}+a_{1}(z)(\hat{a}+\hat{a}^{\dagger})+a_{2}(z)(\hat{a}^{2}+\hat{a}^{\dagger 2})\right]. 	
		\end{equation}
		Using the form of $\hat{a}$ and $\hat{a}^{\dagger}$ in terms of normalized position and momentum operators
		$$\hat{a}=\frac{1}{\sqrt{2}}\left(\hat{X}+i\hat{P}\right)$$
		$$\hat{a}^{\dagger}=\frac{1}{\sqrt{2}}\left(\hat{X}-i\hat{P}\right),$$
		the Hamiltonian becomes,
		\begin{equation}
		\mathcal{H}^{\prime}=-\left[\frac{\hat{P}^{2}}{2 M(z)}+\frac{M(z)\Omega^{2}(z)\hat{X}^{2}}{2}+\sqrt{2}a_{1}(z) \hat{X}-\frac{a_{0}(z)}{2}\right],
		\end{equation}
		where $$M(z)=\frac{1}{a_{0}(z)-2a_{2}(z)}$$ $$\Omega^{2}(z)=a_{0}^{2}(z)-4a_{2}^{2}(z).$$
		We can simplify the problem by considering the solution with a displacement and an overall phase factor as
		$$\vert\Psi(z)\rangle=e^{-i\int \varPhi(z)dz}e^{-i\left[u(z)\hat{P}+M(z)\dot{u}(z)\hat{X}\right]}\vert\psi(z)\rangle,$$
		where the role of first coupling function $a_{1}(z)$ is only by defining the auxillary function $u(z)$ (see ref~\cite{Rodr2014} for complete expressions of $\varPhi(z)$ and $u(z)$). Thus the differential equation (\ref{ShcrLikeEq}) will be modified as
		\begin{equation}
		\left[\frac{\hat{P}^{2}}{2 m(t)}+\frac{m(t)\omega^{2}(t)\hat{X}^{2}}{2}\right]\vert\psi(z)\rangle=i\frac{\partial \vert\psi(z)\rangle}{\partial z},
		\end{equation}
		where $m(t)=M(-z)$ and $\omega(t)=\Omega(-z)$. The above equation expresses the differential equation for a photonic lattice as a mass varying harmonic oscillator Hamiltonian. The same consideration is true for the harmonic oscillator with constant mass. However, it restricts the freedom of arbitrary control over the coupling functions. We consider the usage of STA for this Hamiltonian as we developed in section 4. Unlike the work done by Dionisis Stefanatos~\cite{Stefanatos2014}, this STA method based on Invariant $\mathcal{\hat{I}}^{\prime}$~(\ref{invariantM}) works for any arbitrary variation in mass and we fixed the lattice parameters for the desired output and tried to produce the same output for various propagation distances using the class of invariants given in equation (\ref{invariantM}). To illustrate the control on propagation distance to get the desired output, we can take the example given in ref~\cite{Rodr2014} with parameters, 
		\begin{equation}
		\Omega(z)=\frac{\left[3+\epsilon\tanh(z-z_{s})\right]}{2},~
		a_{0}(z)=\frac{\left[M^{2}(z)\Omega^{2}(z)+1\right]}{2M(z)},~
		a_{1}(z)=1,~
		a_{2}(z)=\frac{\left[M^{2}(z)\Omega^{2}(z)-1\right]}{4M(z)}.
		\label{setParaM}
		\end{equation}

		All the above parameters are plotted in Figure \ref{varparaZ} for $M(z)=1$, where the frequency is a smooth step function, and the constant $\epsilon$ decides the steepness of the curve. We have considered the frequency function for $\epsilon=0.5$, where the initial and final required frequency is tending close to 1 and 2, respectively. This lattice is equivalent to a Glauber-Fock oscillator lattice that makes transitions smoothly from just first-neighbor couplings to first- and second-neighbor couplings~\cite{Rodr2014,Robert2012}. We consider the desired output as the one corresponding to $a_{0}(z_{\tau})=\frac{5}{2}$, $a_{1}(z_{\tau})=1$ and $a_{2}(z_{\tau})=\frac{3}{4}$ while the initial parameters fixed as $a_{0}(0)=a_{1}(0)=1$ and $a_{2}(0)=0$.  Here we have the freedom to decide the arbitrary selection of the mass function during the shortcut process, assigning initial and final values as 1. Considering the equation (\ref{gammaMass}) connecting mass variation and $\Gamma$, the above boundary conditions for mass variations will be in good agreement with the boundary conditions for the specific form of $\Gamma$ of the shortcut process for the harmonic oscillator in section 4. Changing the variables of both the equations (\ref{gammafun}) and (\ref{gammaMass}) in terms of propagation distance, we will obtain similar functions as
		\begin{equation}
		\Gamma(z)=s^{3}(s-1)^{3},
		\label{gammafun1}
		\end{equation}
		and
		\begin{equation}
		\Gamma(z)=\ln M(z),
		\label{gammaMass1}
		\end{equation}
		where $s=\frac{z}{z_{\tau}}$ and $z_{\tau}$ is the location where we need to get the final values of parameters. From the above equations, we get
		\begin{equation}
		M(z)=e^{s^{3}(s-1)^{3}}.
		\label{MEq}
		\end{equation}
		A protocol similar to the shortcut protocol in section 4 will redefine the lattice parameters (index of refraction, first- and second-couplings parameters) through the new propagation distance-dependent functions $\Omega(z)$ and $M(z)$ to control the location of output in the array of the waveguides. A propagation distance-dependent scaling factor similar to the one in equation (\ref{scalefac}),
		\begin{equation}
		\rho(z)=6 \left( \sqrt{\frac{\Omega(0)}{\Omega(z_{\tau})}}-1\right) s^{5}-15\left(\sqrt{\frac{\Omega(0)}{\Omega(z_{\tau})}}-1\right) s^{4}+10\left(\sqrt{\frac{\Omega(0)}{\Omega(z_{\tau})}}-1\right) s^{3}+1,
		\label{scalefac1}
		\end{equation}
		can be used to construct such a protocol. In the above equation $\Omega(0)$ and $\Omega(z_{\tau})$ are the boundary values of frequency $\Omega(z)$. Variation of parameters (index of refraction and coupling parameters) resulting from the set of equations (\ref{setParaM}) is plotted in Figure \ref{varparasta1}-\ref{varparasta10}, and the variation in mass $M(z)$ is plotted in Figure \ref{Massvar} with initial parameters $\Omega(0)=M(0)=1$ and final parameters $\Omega(z_{\tau})=2, M(z_{\tau})=1$. Irrespective of how the parameters vary over the propagation distance, we could drive it from the desired initial to the final values. This mechanism can be used to construct output at necessary locations by arbitrarily controlling the lattice parameters. However, the cost for implementing the protocol is not measurable with the methods explained in section 4 since the working of the photonic lattice is different from that of a single harmonic oscillator.   
		
	\end{section}
	\begin{section}{Conclusion}
		We have successfully derived a class of invariants for the harmonic oscillator under time-dependent frictional force. In the Ermakov equation, the frequency is shifted by the terms with TDCF and its first derivative. The scope of a general approach to STA using the invariant for a QHO with TDCF is studied and found it is feasible, but the specific form of the TDCF decides the boundary conditions for the derivative of scaling factor and the initial and final values of frequency alone defines the boundary values for scaling factor. A compelling case of STA protocol for the quantum harmonic oscillator is framed by identifying a particular class of frictional interaction, such that the $\Gamma(t)$ should be zero at both ends of the shortcut process. We have analyzed the variation of adiabaticity parameter $Q^{*}$, the expectation value of energy $\langle\hat{H}^{IE}\rangle$ and the cost of the shortcut process $\langle\hat{H}^{STA}\rangle$ for various time scales. By interpreting the harmonic oscillator system under time-dependent frictional force as a harmonic oscillator with time-varying mass, we can use the same shortcut protocol to control the dynamics of the harmonic oscillator with inbuilt variation in mass. We have illustrated the coupled photonic lattice case by identifying the propagation of light through the array of waveguides as the evolution of harmonic oscillator wavefunction. The protocol can be improved by formulating some other intelligent scaling factor obeying the corresponding boundary conditions. We have left space for such works with different $\rho(t)$ for improved shortcut protocol characteristics. It is also possible to incorporate such improved shortcut protocols to quantum thermal machines and other applications for enhanced performance. The robustness of the protocol against a possible shift in the frequency of inverse engineered Hamiltonian is analyzed and found it is highly stable. The scope of future research work on the stability analysis of our protocol against various kinds of errors (systematic errors, noise, etc.) can be explored, and it is essential for the experimental implementation of the approach. It is also possible to consider future works that the arbitrary selection of the scaling factor gives more freedom to fix the boundary conditions to optimize the protocol against various errors. The stability analysis studies will also be expected to unveil the ideas to overcome noise and execute STA in a noiseless environment, where we can consider all the possible frequency ranges for QHO. Further, our study might be useful for the applications analogous to the photonic lattice case, which can be studied using harmonic oscillator Hamiltonian with variation in mass.
		
	\end{section}

\end{document}